\documentclass[12pt,preprint]{aastex}

\newcommand{\der}[2][\;\;]{\ensuremath{ \frac{d{#1}}{d{#2}} }}
\newcommand{\dern}[3][\;\;]{\ensuremath{ \frac{d^{#3}{#1}}{d{#2}^{#3}} }}
\newcommand{\dpar}[2][\;\;]{\ensuremath{ \frac{\partial{#1}}{\partial{#2}} }}
\newcommand{\dparn}[3][\;\;]{\ensuremath{ \frac{\partial^{#3}{#1}}{\partial{#2}^{#3}} }}

\newcommand{\bvec}[1]{{\mbox{{\boldmath$#1$}}}}		
\newcommand{\unitv}[1]{\bvec{\hat{#1}}}			
\newcommand{\grad}{\bvec{\nabla}}			
\newcommand{\va}{V_{\rm A}}				

\newcommand{\D}{\displaystyle}
\newcommand{\eqnref}[1]{(\ref{#1})}

\begin{document}

\title{Eigenmodes of 3-D magnetic arcades in the Sun's corona}
\shorttitle{Eigenmodes of 3-D magnetic arcades}

\author{Bradley W. Hindman}
\affil{JILA, NIST and University of Colorado, Boulder, CO~80309-0440, USA}

\author{Rekha Jain}
\affil{School of Mathematics \& Statistics, University of Sheffield, Sheffield S3 7RH, UK}

\email{hindman@solarz.colorado.edu}


\begin{abstract}
We develop a model of coronal-loop oscillations that treats the observed bright
loops as an integral part of a larger 3-D magnetic structure comprised of the entire
magnetic arcade. We demonstrate that magnetic arcades within the solar corona can
trap MHD fast waves in a 3-D waveguide. This is accomplished through the construction
of a cylindrically symmetric model of a magnetic arcade with a potential magnetic
field. For a magnetically dominated plasma, we derive a governing equation for MHD
fast waves and from this equation we show that the magnetic arcade forms a 3-D
waveguide if the Alfv\'en speed increases monotonically beyond a fiducial radius.
Both magnetic pressure and tension act as restoring forces, instead of just tension
as is generally assumed in 1-D models. Since magnetic pressure plays an important
role, the eigenmodes involve propagation both parallel and transverse to the magnetic
field. Using an analytic solution, we derive the specific eigenfrequencies and eigenfunctions
for an arcade possessing a discontinuous density profile. The discontinuity separates
a diffuse cylindrical cavity and an overlying shell of denser plasma that corresponds
to the bright loops. We emphasize that all of the eigenfunctions have a discontinuous
axial velocity at the density interface; hence, the interface can give rise to the
Kelvin-Helmholtz instability. Further, we find that all modes have elliptical
polarization with the degree of polarization changing with height. However, depending
on the line of sight, only one polarization may be clearly visible.
\end{abstract}



\section{Introduction}
\label{sec:Introduction}

The loops of bright plasma revealed in EUV images of the solar corona often sway
back and forth in response to nearby solar flares (e.g., Aschwanden et al. 1999;
Nakariakov et al. 1999). A flurry of observational and theoretical efforts has been
devoted to the explanation and exploitation of these oscillations \citep[see the
review by ][]{Andries:2009}. In particular, the detection of multiple, co-existing
frequencies \citep{Verwichte:2004, VanDoorsselaere:2007, DeMoortel:2007} has encouraged
the hope that seismic methods might be developed to measure loop properties, as first
suggested by \cite{Roberts:1984}. However, before seismic techniques can be fruitfully
applied, we must have a firm grasp on the nature of the wave cavity in which the
waves reside. A commonly accepted viewpoint is that each visible loop is a separate
wave cavity for MHD kink waves. The waves are presumed to have a group velocity that
is parallel to the axis of the magnetic field and each loop oscillates as a coherent,
independent entity. Thus, the problem can be reduced to a 1-D wave problem with boundary
conditions at the two foot points of the loop in the photosphere.

\begin{figure*}
	\epsscale{0.5}
	\plotone{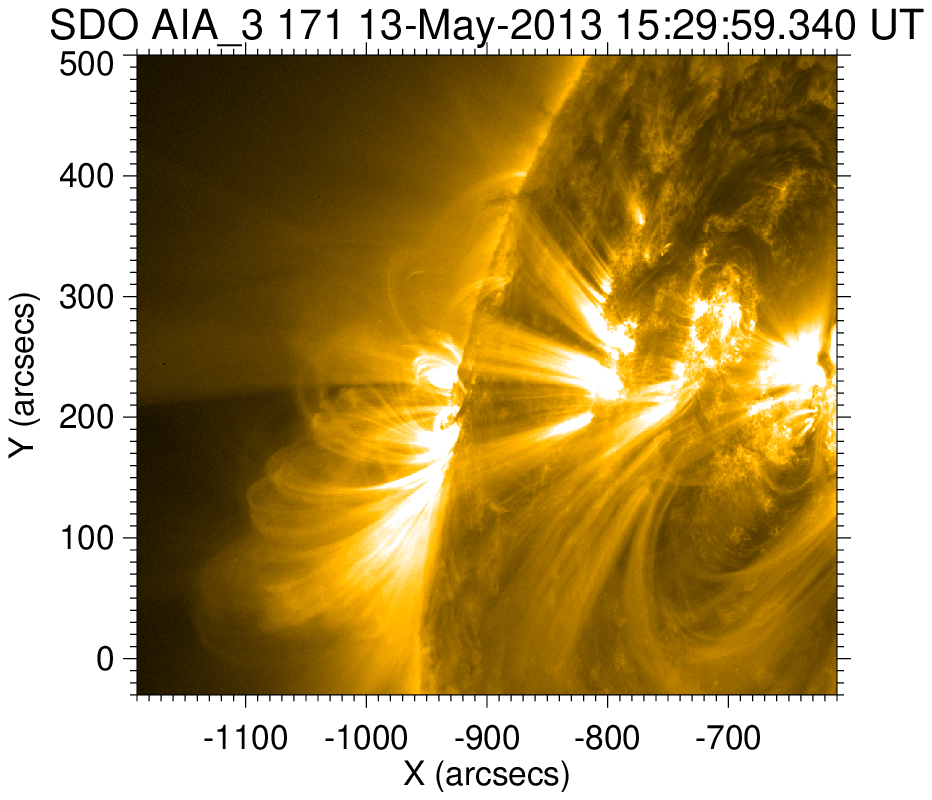}
	\caption{\small Image of an arcade of magnetic field lines in the corona
located on the solar limb. The image was obtained by AIA in the Fe IX 17.1 nm
bandpass. Just after this image was taken, a flare occurred causing the arcade to
spasm. The ancillary file consists of an animation that clearly demonstrates that
the entire arcade participates in the resulting oscillation and that waves propagate
across magnetic field lines.  
	\label{fig:still}}
\end{figure*}

There are several lines of evidence that suggest that the entire 3-D magnetic arcade
in which the bright loops reside participates in the oscillation. Thus, the true
wave cavity is much larger than the individual loop and probably multi-dimensional.
One of these lines of evidence comes from viewing an arcade on the limb where the
loops have higher visibility (see Figure~\ref{fig:still}). With high cadence data
from TRACE and AIA, it is clear that multiple loops within a single magnetic arcade
often oscillate in concert \citep{Verwichte:2009, Jain:2015}. Further, one sees
oscillations traveling down the axis of the arcade, crossing from one loop top to
another, almost perpendicular to  the field lines. To see for yourself, we suggest
that the reader should examine the ancillary file, which consists of an animation
of AIA images that shows the response of a magnetic arcade to a flare. The first
frame of this animation appears is shown separately as Figure~\ref{fig:still}.

A second line of evidence comes from the power spectrum of the oscillations. \cite{Jain:2015}
observed a pair of loops embedded within a common arcade. From the time series of
loop positions, they found that the spectral power was sharply peaked around the
oscillation frequency, but with a noticeable enhancement of the high-frequency wing
of the peak compared to the low-frequency wing. Any scattering process usually results
in symmetric broadening of the power peak. Therefore, \cite{Jain:2015} suggest that
the asymmetry in the power spectrum is evidence for the existence of a 2D or 3D waveguide
with a continuum of eigenmodes propagating transverse to the field as well as parallel.
Each of the modes with transverse propagation has a higher frequency than the mode with
purely parallel propagation. Hence the integral over all possible transverse wavenumbers
results in preferential enhancement of the high-frequency wing of the spectral peak.
Such asymmetry was previously predicted in a theoretical model by \cite{Hindman:2014}
who explored the propagation of fast MHD waves within a 2-D waveguide.

Our goal here is to perform the initial theoretical analysis demonstrating that
magnetic arcades form 3-D waveguides and to elucidate the basic properties of the
resulting 3-D wave modes. In section 2, we present a simple model that treats a
magnetic arcade as a cylindrically symmetric, potential magnetic field. In section 3,
we develop general governing equations for fast MHD waves within the prescribed magnetic
field. In Section 4, we present and discuss an analytic solution for the eigenmodes
for special profiles of the mass density and Alfv\'en speed. Finally, in section 5,
we discuss the observational implications of our work.


\section{Model of a coronal arcade as a magnetized cylinder}
\label{sec:ArcadeModel}

A coronal loop is believed to be bright because preferential heating on a thin
bundle of field lines causes hot, dense plasma to fill that bundle. Therefore, an
arcade of bright loops is probably a magnetic structure where a density inversion
has occurred. A dim cavity of diffuse magnetized plasma underlies a thin region of
dense overlying fluid which radiates profusely, creating the visible loops. Thus,
when modeling a coronal loop, one should attempt to construct a model that possesses
a density enhancement suspended within the corona.

We choose to keep the magnetic field itself relatively simple and consider a 3-D
model which assumes that the magnetic field lines form a set of semi-circular arches.
The magnetic structure possesses cylindrical geometry with the axis of the cylinder
pointing horizontally and embedded in the photospheric plane. The half of the cylinder
lying above the photosphere corresponds to the arcade, with the field lines being
circles centered on the axis. We define a cylindrical coordinate system ($r$, $\theta$, $y$)
whose axis is co-aligned with the axis of the arcade. The axial coordinate is $y$,
the distance from the axis is $r$, and the angle between the position vector and the
photospheric plane is $\theta$. The range of azimuths $\theta \in (0,\pi)$ lies above
the photosphere.

The magnetic field is purely toroidal and force free. The only field that meets these
two conditions is the potential field generated by a line current of strength $I$
located on the $y$ axis,

\begin{equation} \label{eqn:potential_field}
     \bvec{B} = B(r) \, \unitv{\theta} = \frac{2I}{r} \, \unitv{\theta}\; .	
\end{equation}
 
\begin{figure*}
	\epsscale{0.5}
	\plotone{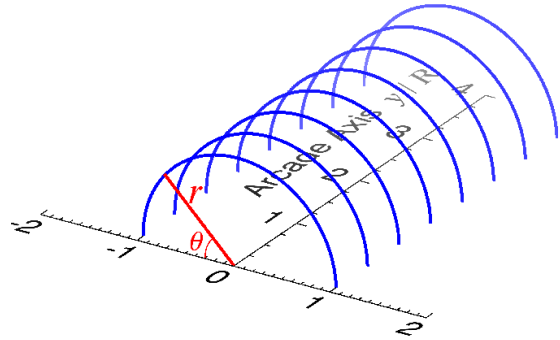}
	\caption{\small Schematic diagram of a 3-D cylindrical coronal arcade. The axis
of the arcade lies embedded in the photosphere and points in the $y$-direction. Field
lines lying within one of the many flux surfaces are indicated by the blue semicircles.
This particular flux surface has a radius of $R$. The arcade lacks shear and is invariant
along its axis. The red ray and arc indicate the radial and angular coordinates of
the cylindrical coordinate system, $r$ and $\theta$ respectively. For the sake of
illustration, all coordinates have been non-dimensionalized by the radius $R$.
	\label{fig:Arcade}}
\end{figure*}

\noindent The field strength $B(r)$ is a decreasing function of radius that only
depends on radius. Figure~\ref{fig:Arcade} illustrates the magnetic geometry.


\section{Governing Wave Equations}
\label{sec:governing_eqns}

If we assume that the plasma is magnetically dominated, so that we can safely ignore
gas pressure and buoyancy forces, the wave motions become purely transverse because
the Lorentz force itself is transverse. Therefore, the azimuthal component, $u_\theta$,
of the fluid's velocity vector will be identically zero, and only the radial and axial
components are nonzero, $\bvec{u} = u_r \, \unitv{r} + u_y \, \unitv{y}$. For such
transverse motions, the induction equation dictates that the fluctuating magnetic
field $\bvec{b}$ is as follows:

\begin{eqnarray}
	\dpar[\bvec{b}]{t} &=& \frac{B}{r} \dpar[u_r]{\theta} \, \unitv{r}
		+ B \Phi \, \unitv{\theta} + \frac{B}{r} \dpar[u_y]{\theta} \, \unitv{y} \; ,
\\
	\Phi &\equiv& -\grad \cdot \bvec{u} + \frac{2u_r}{r} \; .
\end{eqnarray}

\noindent The variable $\Phi$ is proportional to the temporal derivative of the
magnetic-pressure fluctuation, $\Pi$,

\begin{equation}
	\dpar[\Pi]{t} = \dpar[]{t}\left(\frac{\bvec{B} \cdot \bvec{b}}{4 \pi}\right)
		= \frac{B^2}{4 \pi} \Phi \; .
\end{equation}

\noindent For the magnetically dominated plasma discussed previously, the MHD momentum
equation takes on a relatively simple form,

\begin{equation} \label{eqn:vector_MHD}
	\dparn[\bvec{u}]{t}{2}  = \frac{\va^2}{r^2}
		\left(\dparn[u_r]{\theta}{2} \, \unitv{r}
			+ \dparn[u_y]{\theta}{2} \, \unitv{y} \right)
		- \va^2 \grad_\perp \Phi \; ,
\end{equation}

\noindent where $\va$ is the Alfv\'en speed and $\grad_\perp$ is the component of the gradient
operator that is transverse to the magnetic field,

\begin{equation}
	\grad_\perp \equiv \unitv{r} \dpar[]{r} + \unitv{y} \dpar[]{y} \; .
\end{equation}

\noindent In equation~\eqnref{eqn:vector_MHD} the term involving $\Phi$ represents the effects
of magnetic pressure forces and the two terms in parentheses comprise the magnetic
tension.

In order to ensure that equation~\eqnref{eqn:vector_MHD} is tractable and has
separable solutions, we assume that the Alfv\'en speed is a function of radius
alone, $\va=\va(r)$. Since the magnetic-field strength is also a function of only
radius, $B(r) = 2I/r$, the mass density must also vary only with radius,
$\rho = \rho(r)$. This density variation is consistent with hydrostatic balance
along field lines as long as the corona is exceedingly hot such that the density
scale height due to gravitation is much larger than the height of the loops in the
arcade.

For an atmosphere possessing the cylindrical symmetry discussed above, the eigensolutions
to equation~\eqnref{eqn:vector_MHD} have a separable form,

\begin{equation} \label{eqn:eigform}
	\sim \sin\left(m\theta\right) \, e^{iky} \, e^{-i\omega t} \; .
\end{equation}

\noindent In this expression, $\omega$ is the temporal frequency, $k$ is the axial
wavenumber, and $m$ is the azimuthal order. We have selected this solution in order
to satisfy a line-tying boundary condition (i.e., stationary field lines) at the
photosphere ($\theta = 0$ and $\theta = \pi$). Further, we have also assumed that
the arcade is sufficiently long in the axial $y$-direction that we can ignore edge
effects and presume invariance in the $y$ coordinate. Hence, the eigenfunctions are
propagating waves in the $y$-direction. We recognize that this assumption is problematic
for many arcades, but we adopt it despite these reservations for reasons of tractability
and simplicity of argument.

Given solutions with the separable form posited by equation~\eqnref{eqn:eigform}, the
two components of equation~\eqnref{eqn:vector_MHD} become a set of coupled ODEs,

\begin{eqnarray}
	\label{eqn:ur_ODE}
	\left(\omega^2 - \frac{m^2}{r^2}\va^2\right) u_r &=& \va^2 \der[\Phi]{r} \; ,
\\
	\label{eqn:uy_ODE}
	\left(\omega^2 - \frac{m^2}{r^2}\va^2\right) u_y &=& \va^2 ik\Phi \; ,
\\
	\Phi &=& -\der[u_r]{r} + \frac{u_r}{r} - iku_y \; .
\end{eqnarray}

\noindent These equations can be combined into a single ODE for the variable $\Phi$,

\begin{equation} \label{eqn:PhiODE}
	\dern[\Phi]{r}{2} + \left(\frac{1}{r} - \frac{1}{\Lambda}\right) \der[\Phi]{r}
		+\left(\frac{\omega^2}{\va^2} - \frac{m^2}{r^2} - k^2\right) \Phi = 0 \; ,
\end{equation}

\noindent where $\Lambda$ is a scale length

\begin{equation}
	\Lambda^{-1} \equiv \der[]{r}\ln\left(\frac{\omega^2 r^2}{\va^2}-m^2\right) \; .
\end{equation}

Depending on the profile of the Alfv\'en speed as a function of radius, $\va(r)$,
equation~\eqnref{eqn:PhiODE} may possess turning points and critical points. If
turning points exist, there is a possibility that the arcade forms a waveguide.
For example, there might be two turning radii between which the waves are trapped.
In such a circumstance, each eigenmode would be a standing wave in radius $r$ and
azimuth $\theta$, while being a propagating wave in the axial coordinate $y$. The
nature of this type of waveguide will be discussed more fully when we consider
specific Alfv\'en speed profiles in section~\S\ref{subsec:two_shell}.

Potential critical points correspond to cylindrical sheets where the Alfv\'en resonance
condition is satisfied,

\begin{equation}
	\omega^2 = \frac{m^2}{r^2} \va^2(r) \; .
\end{equation}

\noindent Since the scale length $\Lambda$ is divergent at these resonances, the
coefficient of the first-order derivative term in equation~\eqnref{eqn:PhiODE}
becomes singular at these critical radii. Critical layers of this sort often act
as internal reflecting or scattering interfaces.  However, since our subsequent arcade
models will be intentionally constructed such that no critical radii exist, we
defer further exploration of their behavior.


\section{Analytic Solution}
\label{sec:analytic_sol}

In order to provide a specific example of the waveguides that can form within these
magnetic arcades, we present an analytic solution to equation~\eqnref{eqn:PhiODE}.
First, we note that if the waves are to be trapped in the radial direction, the
Alfv\'en speed must increase monotonically beyond a fiducial radius. This is necessary
so that refraction occurs and turns outward propagating waves back toward the interior.
This places a restriction on the density profile of any model atmosphere that one
proposes.  Since, the magnetic-field strength decreases as $1/r$, in order for the
Alfv\'en speed to increase with radius, the density must decrease with radius faster
than $1/r^2$.

The analytic solution that we present here is predicated on the removal of all critical
radii and is accomplished by specifying a particular profile for the Alfv\'en speed.
Consider an Alfv\'en speed profile that increases linearly with cylindrical radius
(thus satisfying the condition required for a refractive turning point). For such a
profile, the Alfv\'en resonances disappear, the density decreases with radius as
$1/r^4$, and the scale length $\Lambda$ disappears from equation~\eqnref{eqn:PhiODE},

\begin{eqnarray}
	\va(r) &=& V \left(\frac{r}{r_0}\right) \; ,
\\
	\rho(r) &=& \frac{I^2}{\pi r_0^2 V^2} \left(\frac{r_0}{r}\right)^4 \; ,
\\
	\Lambda^{-1} &=& 0 \; .
\end{eqnarray}

In these expressions, $V$ and $r_0$ are constant reference values that provide the
constant of proportionality such that $\va = V$ at radius $r = r_0$. Since the
reciprocal of $\Lambda$ vanishes, the ODE describing the radial behavior of the
eigenfunctions can be recognized as a modified Bessel function equation,

\begin{equation} \label{eqn:AnalODE}
	\dern[\Phi]{r}{2} + \frac{1}{r} \der[\Phi]{r}
		-\left(k^2 + \frac{\nu^2}{r^2}\right) \Phi = 0 \; ,
\end{equation}

\noindent where the order $\nu$ is potentially imaginary for sufficiently high frequency
$\omega$,

\begin{equation}
	\nu^2 = m^2 - \frac{\omega^2 r_0^2}{V^2} \; .
\end{equation}

\noindent The two linearly independent solutions for $\Phi$ are just the modified Bessel
functions $I_\nu(kr)$ and $K_\nu(kr)$. The velocity components can be derived from the
dimensionless pressure fluctuation, $\Phi$, by using equations~\eqnref{eqn:ur_ODE} and
\eqnref{eqn:uy_ODE},

\begin{eqnarray}
	u_r &=& -\frac{r^2}{\nu^2} \frac{d\Phi}{dr} \; ,
\\
	u_y &=& -\frac{r^2}{\nu^2} ik\Phi \; .
\end{eqnarray}


\subsection{Nature of the Solutions}
\label{subsec:nature_sol}

The order $\nu$ of the modified Bessel functions can be either purely real or purely
imaginary and there is a transition frequency that marks the change between real
and imaginary values. For low frequencies, $\omega < m V/r_0$, the order $\nu$ is
purely real, while for high frequencies, $\omega > m V/r_0$, the order is purely
imaginary. For real orders, the $K$ modified Bessel function is the solution that
remains finite as $r\to\infty$, whereas the $I$ modified Bessel function diverges.
At the origin, the $I$ Bessel function remains finite and the $K$ Bessel function
diverges. However, for imaginary order, the $K$ Bessel function is a real function,
it vanishes at infinity, and is finite but recessive at the origin. The recessive
behavior (i.e., oscillatory with an infinite number of zeros on the approach to the
origin) arises because the wave speed vanishes at the accumulation point. The
$I$ Bessel function is a complex function when the order is imaginary and is rather
badly behaved. If needed, an additional real solution can be constructed by taking
a specific linear combination of $I$ modified Bessel functions. This new real function
is known as the $L$ modified Bessel function \citep{Dunster:1990},

\begin{equation}
	L_\nu(z)=\frac{i\pi}{2\sin\left(\nu\pi\right)} \left[ I_\nu(z) + I_{-\nu}(z)\right]\; .
\end{equation}

\noindent For imaginary order $\nu$, the $L$ Bessel function has the properties that
it is divergent at infinity but remains finite at the origin; although like the $K$
Bessel function, it is recessive with an infinite number of zeros piling up at the
origin. We note that neither of these solutions is particularly realistic because
of the recessive behavior at the origin. However, if we had permitted the Alfv\'en
speed to remain nonzero at the origin, these difficulties would have been automatically
avoided. In section~\S\ref{subsec:two_shell}, we solve this problem in another manner
by constructing piece-wise continuous models of the atmosphere such that the waves are
excluded from the origin by enforcing their evanescence in the inner shell.

The behavior of these solutions can be predicted and more fully understood by deriving
a local radial wavenumber $\kappa_r$ for equation~\eqnref{eqn:AnalODE}. We do this by
performing a change of variable, $\Psi = r^{1/2} \Phi$, such that the transformed equation
becomes a Helmholtz equation lacking a first derivative term,

\begin{equation} \label{eqn:PhiWKB}
	\dern[\Psi]{r}{2} - \left(\frac{\nu^2-1/4}{r^2} + k^2\right) \Psi = 0 \; .
\end{equation}

\noindent The local radial wavenumber can be read-off from this equation and, if expressed
in terms of frequency $\omega$ instead of $\nu$, we find

\begin{eqnarray}
	\label{eqn:loc_disp}
	\kappa_r^2 &=& \frac{\omega^2-\omega_c^2}{\va^2} - k^2 \; ,
\\
	\label{eqn:cut-off}
	\omega_c^2 &\equiv& \left(m^2-1/4\right)\frac{V^2}{r_0^2} \; .
\end{eqnarray}

\begin{figure*}
	\epsscale{0.5}
	\plotone{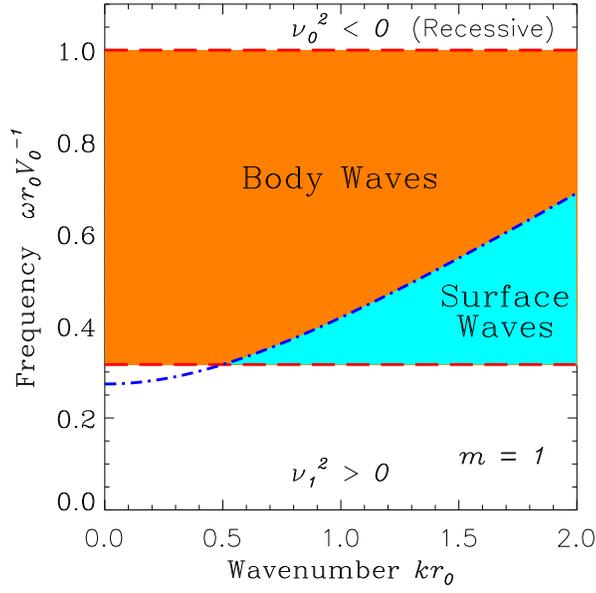}
	\caption{\small A propagation diagram showing the parameter regime in which
the allowed body-wave and surface-wave solutions exist. The two red dashed lines 
indicate the upper and lower limits given by equations~\eqnref{eqn:upper_limit}
and \eqnref{eqn:lower_limit}. The blue dotted-dashed curve indicates the boundary
between the body and surface waves, i.e., equation~\eqnref{eqn:body_surf}. Above
the upper limit, the solutions become recessive at the origin.  Below the lower limit,
it can be shown that the transcendental dispersion relation~\eqnref{eqn:disp_relation}
has no solutions that satisfy the necessary continuity conditions at the interface
between the inner the outer shells.
	\label{fig:Prop_diag}}
\end{figure*}

\noindent This local dispersion relation has been written in terms of a spatially constant
($m$-dependent) cut-off frequency, $\omega_c$. High frequency waves,
$\omega^2 > \omega_c^2 + k^2 \va^2$, propagate, while low frequency waves,
$\omega^2 < \omega_c^2 + k^2 \va^2$, are evanescent. Figure~\ref{fig:Prop_diag} provides
a propagation diagram that illustrates those combinations of frequency $\omega$ and axial
wavenumber $k$ that correspond to propagating waves and those associated with evanescent
waves.

From the local wavenumber we can immediately deduce that there can only be one
turning point and it is located at

\begin{equation} \label{eqn:rturn}
	r_{\rm turn} = \frac{\sqrt{1/4-\nu^2}}{k} = \frac{\sqrt{\omega^2-\omega_c^2}}{k}\; \frac{r_0}{V}  \; .
\end{equation}

\noindent Thus, we only have oscillatory solutions if $\omega^2 > \omega_c^2$ (or
equivalently $\nu^2 < 1/4$). The extent of the waveguide, or the regime of propagation,
will span the range $r \in [0, r_{\rm turn}]$, because there is only one turning point. We
can further deduce that the solutions will be recessive at the origin by noting that
the local radial wavenumber is divergent as $r\to 0$ (i.e., the local wavelength vanishes
at the origin).


\subsection{A Waveguide Comprised of Two Shells}
\label{subsec:two_shell}

Since our goal here is to examine loop models that consist of a density enhancement
suspended above a diffuse cavity, we choose to construct a piece-wise continuous
model that is comprised of two cylindrical shells joined at $r=r_0$. Each shell
possesses the Alfv\'en speed profile that permits the analytic solution, but there
is a pycnocline at the cylindrical interface between the two shells. This interface
is currentless such that the magnetic field is continuous and unmodified from
equation~\eqnref{eqn:potential_field}. However, the Alfv\'en speed $\va$ and mass
density $\rho$ are discontinuous at the interface and given by

\begin{eqnarray}
	\va^2(r) &=& \left\{ \begin{array}{rl}
					\D V_0^2 \left(\frac{r}{r_0}\right)^2 & {\rm if } \, r < r_0 \; , \\
					\D V_1^2 \left(\frac{r}{r_0}\right)^2 & {\rm if } \, r > r_0 \; ,
				\end{array} \right.
\\
	\rho(r) &=& \left\{ \begin{array}{rl}
					\D \frac{I^2}{\pi r_0^2 V_0^2} \left(\frac{r_0}{r}\right)^4 & {\rm if } \, r < r_0 \; , \\
					\D \frac{I^2}{\pi r_0^2 V_1^2} \left(\frac{r_0}{r}\right)^4 & {\rm if } \, r > r_0 \; ,
				\end{array} \right.
\end{eqnarray}

\noindent where $V_0$ is the reference speed in the inner region ($r<r_0$) and $V_1$
is the reference speed in the outer region ($r>r_0$).

Since the Alfv\'en speed drops as $r$ increases across the interface (i.e., $V_1 < V_0$),
the cut-off frequency, $\omega_c$, is larger in the inner shell than it is in the
outer shell---see equation~\eqnref{eqn:cut-off}. This has the salutary property that
there exists a band of frequencies for which the solutions are evanescent in the inner
shell and potentially propagating in regions of the outer shell. Thus, we have the
possibility of trapped waves that lack recessive behavior at the origin. Since the
cut-off frequencies differ in the two regions, the order of the Bessel functions in
the inner region, $\nu_0$, and in the outer region, $\nu_1$, also differ,
$\nu_j^2 = m^2 - \omega^2 r_0^2 / V_j^2$.

A transcendental dispersion relation can be derived for this model by requiring the
continuity of both the magnetic pressure and the radial velocity across the interface
at $r=r_0$. However, we must first choose the proper solutions within the two regions
such that boundary conditions at $r\to\infty$ and $r=0$ are satisfied. In order to
avoid divergent or recessive solutions at the origin, the solution within
the inner region must correspond to an $I$ modified Bessel function and the order
$\nu_0$ in that region must be purely real (i.e., $\nu_0^2 > 0$). This places an upper
bound on the mode frequencies,

\begin{equation} \label{eqn:upper_limit}
	\omega^2 < \frac{m^2}{r_0^2} V_0^2 \; .
\end{equation}

\noindent Similarly, a boundary condition of finiteness as $r\to\infty$ selects
the proper solution in the outer region. The only solution that remains finite in
this limit is the $K$ modified Bessel function and it does so for either real or
imaginary order $\nu_1$.

Given that the solution must be an $I_{\nu_0}(kr)$ function for $r < r_0$ and a $K_{\nu_1}(kr)$
function for $r>r_0$, the continuity of magnetic pressure and radial velocity mandate
the following dispersion relation,

\begin{equation} \label{eqn:disp_relation}
	\nu_0^2 \frac{K_{\nu_1}^\prime(k r_0)}{K_{\nu_1}(k r_0)} = \nu_1^2 \frac{I_{\nu_0}^\prime(k r_0)}{I_{\nu_0}(k r_0)} \; ,
\end{equation}

\noindent and eigenfunction

\begin{equation} \label{eqn:Eigenfunction}
	\Phi(r) = \left\{
	   \begin{array}{rl}
		\D A \, K_{\nu_1}(k r_0) \, I_{\nu_0}(k r) & {\rm if } \, r < r_0 \; , \\
		\D A \, I_{\nu_0}(k r_0) \, K_{\nu_1}(k r) & {\rm if } \, r > r_0 \; ,
	   \end{array} \right.
\end{equation}

\noindent where the primes indicate differentiation with respect to the argument of
the Bessel function and $A$ is a normalization constant.

If one assumes a priori that $\nu_1^2 \geq 0$ one can quickly demonstrate that the left-hand
side of equation~\eqnref{eqn:disp_relation} is negative and the right-hand side is positive.
Hence, a contradiction occurs and only solutions with $\nu_1^2 < 0$ are allowed. This
places a further restriction on the mode frequencies, imposing a lower limit,

\begin{equation} \label{eqn:lower_limit}
	\omega^2 > \frac{m^2}{r_0^2} V_1^2 \; .
\end{equation}

\noindent Thus, the eigenfrequencies exist within a band, namely $m V_1/r_0 < \omega < m V_0/r_0$,
whose extent depends on the wave speeds within each region, $V_0$ and $V_1$. Clearly,
for the band to exist, we must have the ordering $V_1 < V_0$; so, the Alfv\'en speed
must decrease across the interface. This implies that the mass density must increase
across the interface and modes only exist when the inner region forms a diffuse cavity
with denser plasma overlaid.   In all subsequent figures, we adopt a value of
$V_1^2 / V_0^2 = 0.1$, corresponding to a tenfold jump in density across the interface.

The waves can be further decomposed into body waves and surface waves based on their
propagation properties. Body waves will possess a zone of propagation just above the
interface, whereas surface waves will be evanescent throughout both shells. If
$r_{\rm turn} > r_0$ the mode will be a body wave and, conversely, if $r_{\rm turn} < r_0$
the mode corresponds to a surface wave. The demarcation can be expressed in terms of
frequencies and wavenumbers through the use of equation~\eqnref{eqn:rturn}, 

\begin{equation} \label{eqn:body_surf}
	\omega_t^2 = \left(m^2 - 1/4 +k^2 r_0^2\right) \frac{V_1^2}{r_0^2} \; .
\end{equation}

\noindent Waves with $\omega>\omega_t$ are body waves and those with $\omega<\omega_t$
are surface waves.

Figure~\ref{fig:Prop_diag} illustrates the regimes of allowed solutions and the
zones of propagation and evanescence in a propagation diagram for $m=1$.  Other $m$ have similar
but not identical diagrams.  The orange region corresponds to body waves and the turquoise
region to surface waves. The white regions of the diagram correspond to either recessive waves
(at high frequency) or a zone of nonresonant oscillations (at low frequency).

\begin{figure*}
	\epsscale{0.9}
	\plotone{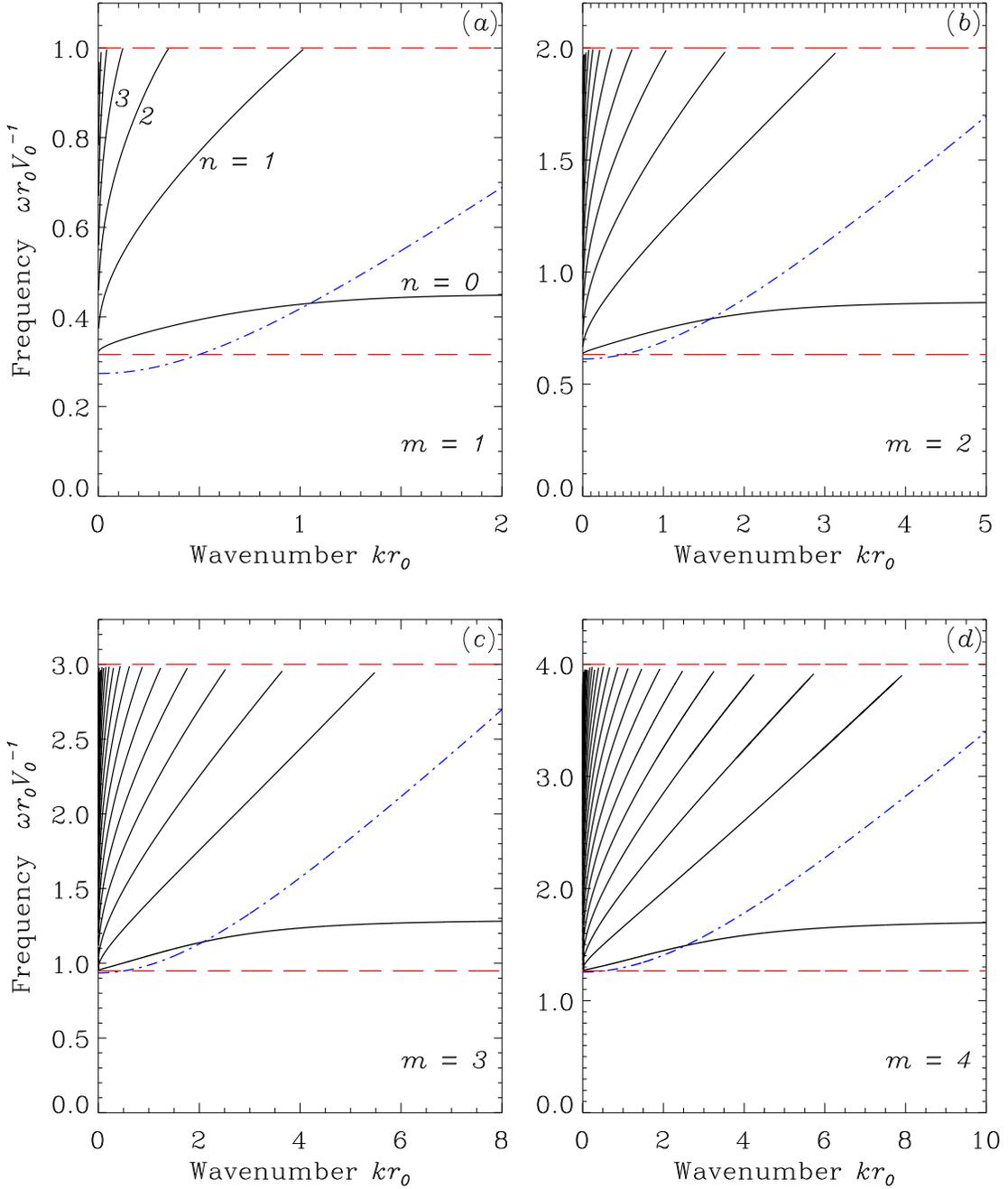}
	\caption{\small Dispersion curves showing the dimensionless frequency $\omega r_0/V_0$
as a function of the dimensionless axial wavenumber $kr_0$. For illustrative purposes,
we have set the density contrast across the density discontinuity to be 10; thus,
$V_1^2 / V_0^2 = 0.1$. The four panels correspond to different axial orders ($m \in[1,4]$)
as indicated in the lower right corner of each panel. The lowest frequency curve in
each panel corresponds the fundamental radial mode $n=0$, and subsequently higher curves
are for $n=1$, $n=2$, and so on. The red and blue curves are the same as in
Figure~\ref{fig:Prop_diag}. All dispersion curves begin near the lower frequency limit
for $k = 0$. The overtones increase in frequency as $k$ increases and disappear over the
upper limit marking the boundary of high-frequency recessive solutions.  Because of this
disappearance, not all radial mode orders exist for any given wavenumber $k$. The fundamental
radial mode ($n=0$) never crosses the upper limit and instead asymptotes to a constant frequency.
	\label{fig:Eigenfrequencies}}
\end{figure*}

Since the dispersion relation is transcendental, we must solve it numerically if
we wish to calculate the mode frequencies as a function of axial wavenumber $k$
and azimuthal order $m$. We use the numerical codes developed by \cite{Gil:2004}
to numerically compute the $K$ Bessel functions of imaginary order and \cite{Press:2007}
to evaluate the $I$ Bessel functions. Figure~\ref{fig:Eigenfrequencies} illustrates
dispersion curves for the fundamental mode ($m=1$) and the first few azimuthal
overtones ($m>1$). Naturally, the solutions possess an integer number of nodes
with radius, and the number of nodes (or the radial order $n$) labels each family
of solutions. Each curve in Figure~\ref{fig:Eigenfrequencies} corresponds to a mode
of specific radial order, $n \in [0,7]$, but different axial wavenumber $k$, which
is of course continuous. Note, as a function of $k$, each modal curve begins near
the cut-off frequency of the outer shell. The radial overtones $n > 0$ continue to
increase with wavenumber until they disappear as they cross the high-frequency limit
demarking recessive behavior at the origin. The gravest mode ($n=0$), on the other
hand, rapidly flattens as $k$ increases and asymptotes to a fixed value.

\begin{figure*}
	\epsscale{0.8}
	\plotone{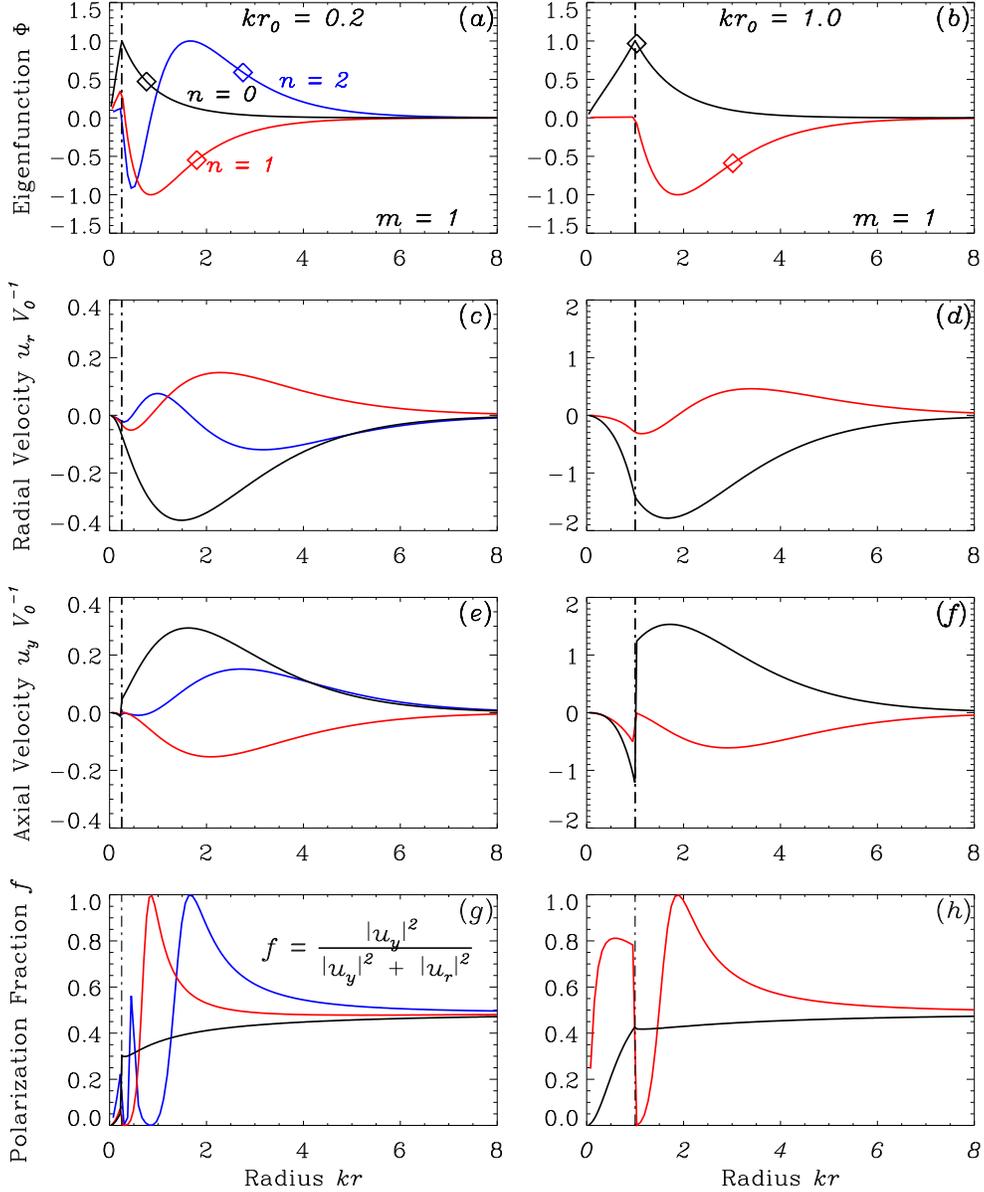}
	\caption{\small Eigenfunctions associated with the eigenvalues presented in
Figure~\ref{fig:Eigenfrequencies}. Only the fundamental azimuthal mode $m=1$ is shown
and results for two different axial wavenumbers are illustrated; the left column
shows $kr_0 = 0.2$ and the right column shows $kr_0 = 1.0$. The first three rows of panels
show dimensionless eigenfunctions as solid curves for the magnetic pressure fluctuation
$\Phi$, the radial velocity $u_r / V_0$, and the axial velocity $u_y / V_0$. The
different colors indicate the radial order of the mode: black ($n=0$), red ($n=1$),
and blue ($n=2$).  Note that not all radial orders exist for all wavenumbers. The vertical
line indicates the location of the interface between the inner and outer shells.
The magnetic pressure fluctuation and the radial velocity are continuous across this
matching layer, whereas the axial velocity is discontinuous and changes sign. The
diamond appearing on each of the curves for $\Phi$ indicate the location of the turning
point $r_{\rm turn}$ for that mode. The final row of panels presents the polarization
fraction $\left|u_y\right|^2 / \left( \left|u_y\right|^2 + \left|u_r\right|^2\right)$.
All waves possess elliptical polarization, with elliptical fluid-parcel orbits. A
value of 1 for the polarization fraction indicates purely axial (or horizontal)
motion, while a value of 0 shows pure radial (or vertical) motion. A value of 0.5
indicates circular polarization.
	\label{fig:Eigenfunctions}}
\end{figure*}

Figures~\ref{fig:Eigenfunctions}$a$--$f$ present radial eigenfunctions for the
dimensionless-pressure fluctuation $\Phi$ (Figures~\ref{fig:Eigenfunctions}$a$ and $b$),
the radial velocity $u_r$ ($c$ and $d$), and the axial velocity $u_y$ ($e$ and $f$).
All of the eigenfunctions are only for $m=1$, although the higher azimuthal orders
have similar behavior. The normalization constant $A$ in equation~\eqnref{eqn:Eigenfunction}
has been chosen purely for illustrative purposes. Each of the panels corresponding to
an axial wavenumber of $kr_0 = 0.2$ show the gravest three radial orders. While for
$kr_0 = 1.0$, only the eigenfunctions for $n=0$ and $n=1$ are shown, because $n=2$
does not exist. The matching conditions ensure that $\Phi$ and $u_r$ are continuous
across the interface between the two shells. The axial velocity $u_y$ is discontinuous
and in fact changes sign across the interface.

In the inner shell, the eigenfunctions correspond to evanescent solutions that vanish
at the origin. Within the outer shell, the solution may be propagating just above the
interface and evanescent higher (if a body wave) or it may be evanescent throughout
(if a surface wave). The lower boundary of the waveguide is the same for all modes and 
it is located at the interface between inner and outer shells. For the body waves, the
upper boundary of the waveguide is different for each mode and corresponds to the turning
point located at $r_{\rm turn} = \sqrt{\omega^2-\omega_c^2}\; (r_0/kV)$. Therefore, the
radial size of the waveguide depends on frequency $\omega$, axial wavenumber $k$, and
azimuthal order $m$ (because $\omega_c$ depends on $m$). These turning points are indicated
in Figures~\ref{fig:Eigenfunctions}$(a)$ and $(b)$ by the colored diamonds. Note that
the turning points should not correspond to the inflection points of $\Phi$. Instead,
they are located at the inflection points of $\Psi = r^{1/2} \Phi$.

The gravest radial mode ($n=0$) is slightly unusual. At low axial wavenumber, the wave
is a body wave with a narrow zone of propagation just above the interface. As the
wavenumber increases, the mode changes into a surface wave that is evanescent everywhere.
Since this mode only lives on the interface, its behavior is quite distinct from the
higher overtones, and hence its dispersion relation does not follow the same pattern
as the radial overtones.


\section{Discussion}
\label{sec:discussion}

We have developed general equations describing the fast MHD eigenmodes of a coronal
magnetic arcade possessing cylindrical geometry and an Alfv\'en speed that is a
function of only the cylindrical radius, $\va(r)$.  We find that a waveguide exists
as long as the Alfv\'en speed monotonically increases beyond some radius. This waveguide
traps waves radially and azimuthally, and allows free propagation of the waves down
the cylindrical axis. Thus, the eigenfunctions of the magnetic arcade consist of
standing waves in radius and azimuth and traveling waves in the axial direction.
We subsequently derived the eigenfunctions and eigenfrequencies for an arcade model with
a specific Alfv\'en speed profile. In this model, the arcade has a diffuse, evacuated
cavity that extends to a cylindrical radius of $r_0$. The edge of the cavity consists
of a density discontinuity over which hangs a shell of denser fluid. Separately in
each region, the Alfv\'en speed is linearly proportional to the radius. Fast MHD waves
can be trapped as body waves that live in the denser region above the density discontinuity
at $r=r_0$ or as surface waves that reside on the density discontinuity itself.

We reiterate that the resonant oscillations of the arcade have both magnetic pressure
and tension as restoring forces. This can be visually verified by noting in
Figure~\ref{fig:Eigenfunctions} that the dimensionless magnetic pressure, $\Phi$,
is always comparable in magnitude to the velocity components, which are themselves
proportional to the magnetic tension. Only waves with zero axial wavenumber $k$ can
be pure tension waves. This is a very different result from what is predicted by 1-D
models that treat the loop as a single coherent entity. Those studies argue that the
oscillations are fast MHD kink waves primarily driven by tension forces.


\subsection{Observational Implications}

Occasionally, the arcade itself may not be fully visible and only one or two preferentially
illuminated loops can be seen. In such cases, fast MHD waves may be traversing the
entire arcade but the 3-D wave field can only be sampled along a given field line.
Since the wave modes of the arcade are trapped, standing waves in the direction parallel
to the field (i.e., reflection from the two photospheric foot points), the motion
of any individual field line may look very reminiscent of a 1-D wave cavity. Only
by careful examination of the spectral content of the oscillations can the existence
of the full 3-D wave field be surmised.

If the wave excitation region is sufficiently broad, we anticipate that the
lowest-order modes in both radius $n=0$ and azimuth $m=1$ will be preferentially
excited. Thus the surface wave is of particular importance. Furthermore, we expect
that most of the emission, which is visible as a bright loop, should arise from the
densest fluid. Hence, the cavity should be dim and the bright loop probably corresponds
to only the narrow portion of the waveguide immediately above the interface at $r=r_0$.
If such a supposition is correct, the radial shear that appears in the axial velocity
may not be evident. Further, even the radial overtones, which have nodes in radius,
may appear as a single collimated oscillating loop.

Finally, we emphasize that each radial order actually comprises a continuum of
eigenmodes with different axial wavenumbers $k$. The relative amplitude of these
modes will depend explicitly on the manner in which the waves are excited. However,
in driven problems the mode with the lowest frequency is usually the most easily
excited. From the dispersion curves, Figure~\ref{fig:Eigenfrequencies}, one would
therefore expect that those modes with $k\approx~0$ would dominate the spectrum,
forming a distinct peak. Modes with nonzero wavenumber have higher frequency and
thus would contribute power exclusively to the high-frequency wing of the peak. The
resulting power spectrum would possess an asymmetric power peak with a deficit of
power in the low-frequency wing. Scattering processes certainly broaden a spectral
peak, but they typically do so symmetrically. Therefore, the existence of asymmetric
power---as observed by \cite{Jain:2015}---may be evidence for transverse wave
propagation within a waveguide.

The surface wave could very well appear in power spectra with a doubly peaked
profile. One peak would appear near the lowest frequency available, associated
with the waves with $k\approx0$. The second, higher-frequency peak would correspond
to an accumulation of power from all of the modes with $k>>1/r_0$ where the dispersion
curve becomes flat. The width of each peak and relative amplitude would of course
depend on the details of the driving.


\subsection{Velocity Polarization}

Coronal loops are observed to oscillate in two polarizations of motion. ``Horizontal"
oscillations involve swaying of the loops back and forth and correspond to axial
motions $u_y$ in the geometry presented here.  ``Vertical" oscillations \citep{Wang:2004} 
cause an expansion and contraction of the loop in radius and in our geometry such
motions $u_r$ would be purely in the radial direction. For many of the observed
loops, only one polarization is actually observable due to projection effects and the
vantage of observation. Hence, one should be mindful that observations of a particular
polarization are only proof that one of the possible polarizations exists, not as
evidence that the other polarization is absent.

The arcade modes presented here never oscillate in a pure polarization, nor do all
of the field lines passing through the waveguide oscillate as a coherent bundle. Instead,
the cross-sectional shape of a sheaf of field lines shears and distorts during a
period of the oscillation. Equations~\eqnref{eqn:ur_ODE} and \eqnref{eqn:uy_ODE}
reveal that the two polarizations are coupled in the 3-D arcade. The eigenfunctions
illustrated in Figure~\ref{fig:Eigenfunctions} demonstrate the coupling by clearly
showing that over most of the parameter regime, the two velocity components have
similar magnitudes. From equations~\eqnref{eqn:ur_ODE} and \eqnref{eqn:uy_ODE}, we
can immediately deduce that the axial and radial motions are 90$^\circ$ out of phase.
Thus, the polarization is actually elliptical and any given parcel of fluid undergoes
elliptical motion in the plane perpendicular to the local field line. The orientation
of the minor and major axes is always the same (axial and radial), but the ellipticity
depends explicitly on the polarization fraction,
$f=\left|u_y\right|^2 / \left(\left|u_y\right|^2+\left|u_r\right|^2\right)$.
This polarization fraction is presented in Figures~\ref{fig:Eigenfunctions}$g$ and $h$
for the same eigenfunctions that we provided earlier. A value of 1 corresponds to pure
axial or horizontal motion and a polarization fraction of 0 indicates pure radial motion
or vertical polarization. A polarization fraction of 0.5 indicates circular polarization.
 
Even the fundamental mode in radius, $n=0$, which has the simplest radial behavior,
changes polarizations as a function of height.  Near the origin, where the mode has
very weak motions, the polarization is primarily radial (or vertical). Throughout
the inner cavity, the polarization shifts until it reaches an even admixture of the two
polarizations within the outer shell (i.e., the polarization is nearly circular). For
higher-order radial modes, the polarization swings back and forth between primarily
axial (horizontal) to primarily radial (vertical), as one passes through the nodes of
the respective eigenfunctions. Above the turning point, in the evanescent region of the
eigenfunction, the polarization becomes nearly circular. However, since we do not expect
the entire waveguide to be visible, the observed motion of a bright coronal loop is
likely to possess a single polarization corresponding to the portion of the eigenfunction
that lies directly above the interface at $r = r_0$. If the fundamental radial mode is
observed, this polarization should be nearly circular (see Figure~\ref{fig:Eigenfunctions}),
but if higher radial orders are observed, the polarization could be predominately radial.


\subsection{Radial Shear in the Axial Velocity}

The interface between the diffuse interior cavity and the outer shell of
dense fluid is a region of intense shear. While the magnetic pressure and radial
velocity are continuous across this layer, the axial velocity changes sign across
the interface. Therefore, even for modes of the lowest radial order, the horizontal
motion within the cavity is in the opposite direction compared to the motions in
the overlying fluid. As stated previously, we suspect that much of this shear may
remain invisible in a real arcade, as the emission generating the image of the bright
loop likely arises from the dense fluid within the outer shell above the shear zone.

Since the interface between the cavity and outer shell is both a region of sharp
increase in the mass density with radius and a zone of strong shear in the horizontal
flow speed, this interface is ripe for the operation of the Kelvin-Helmholtz instability.
Since the magnetic field points in a direction perpendicular to the shear, the instability
is largely unaffected by the presence of the field. The linear growth rate arising
from just the velocity shear (ignoring the effects of an unstable density gradient)
is given by

\begin{equation}
	\gamma = k \frac{\sqrt{V_0^2 V_1^2}}{V_0^2 + V_1^2} \Delta u_y \; ,
\end{equation}

\noindent where $\Delta u_y$ is the difference in the axial velocity across the
interface at $r=r_0$. To estimate a typical value for this growth rate, we use
a transverse loop velocity of 100 m s$^{-1}$ \citep{Jain:2015} and adopt a velocity
jump of twice this value. If we assume that the density increases roughly by a
factor of four across the interface, as is typical for a prominence cavity
\citep{Schmit:2011}, the ratio of square Alfv\'en speeds $V_0^2 / V_1^2$
has the same value.  Thus, we estimate the growth rate to be $\gamma/k = 10^2$ m s$^{-1}$.
A ratio of this number to the local Alfv\'en speed should be constructed to determine
the magnitude of the growth rate compared to a fast wave frequency. Given an
Alfv\'en speed of 2000 km s$^{-1}$. We estimate that the ratio of the growth rate
to wave frequency is on the order of $10^{-4}$. Therefore, the growth rate is likely
to be quite weak and any turbulent flows arising from the shear in the eigenfunctions
are also likely to be small perturbations.

      
\subsection{Surface Waves}

Since the modes of lowest radial and azimuthal order are more efficiently excited
by large-scale disturbances, we expect the $n=0$ mode to predominate. For low axial
wavenumber $k$ this mode is a body wave.  For large $k$ the mode transitions to
a pure surface wave. In either case, the magnetic pressure fluctuation and the
radial velocity have single lobed eigenfunctions.  The magnetic pressure peaks at
the interface at $r = r_0$ and the radial velocity peaks rather higher. The axial
velocity is discontinuous at the interface and changes sign across it, but otherwise
lacks zeros.

\begin{figure*}
	\epsscale{0.5}
	\plotone{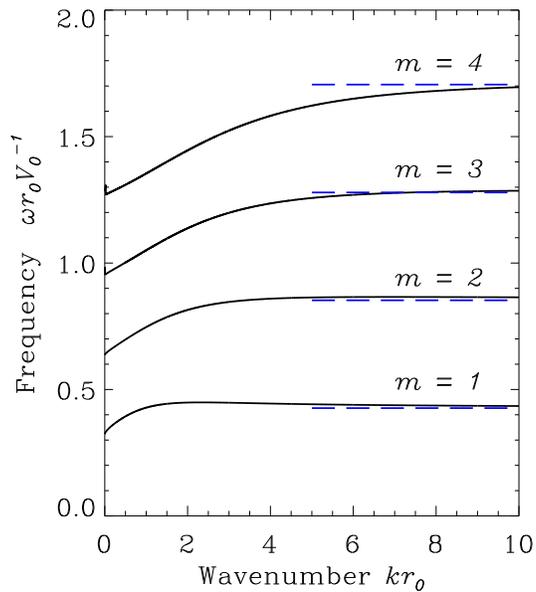}
	\caption{\small Dimensionless eigenfrequencies of surface waves as a function
of dimensionless  axial wavenumber $kr_0$. We present dispersion curves for the four
lowest azimuthal orders $m \in [1,4]$. In the limit of infinite wavenumber, all frequencies
approach an $m$-dependent asymptotic value.  Very careful examination of the curve for
$m=1$ reveals that the curve actually achieves a shallow maximum near a wavenumber of
$kr_0 = 2$. Thus, the approach to the asymptote is not necessarily monotonic. The asymptotic
value estimated by equation~\eqnref{eqn:asymptotic} is demarked by the blue dashed lines.
	\label{fig:Surf_waves}}
\end{figure*}

The dispersion relation for the fundamental radial mode is rather different from
that of the overtones. First, the $n=0$ mode exists for all wavenumbers. This occurs
because the eigenfrequency does not continue to rise as the wavenumber increases and
thus does not disappear as it crosses over the upper frequency limit. In fact, the
dispersion curve for the surface wave asymptotes to a finite value as the wavenumber
increases to infinity. Figure~\ref{fig:Surf_waves} illustrates the behavior of the
dispersion curves for $m \in [1,4]$. The curve corresponding to $m=1$ demonstrates
that the approach to the asymptotic value is not always monotonic.

A quick asymptotic analysis of equation~\eqnref{eqn:PhiWKB} for large axial
wavenumbers $kr_0 \to \infty$ confirms the results achieved by numerical evaluation
of the full solution, namely, that for the two-shell model the surface wave has a
frequency that becomes independent of $k$, approaching a constant value that depends
on the wavenumber parallel to the field lines $m/r_0$ and to the fractional change
in the Alfv\'en speed across the interface $V_1/V_0$, 

\begin{equation} \label{eqn:asymptotic}
	\omega^2 \approx \frac{2m^2}{r_0^2} \frac{V_0^2 V_1^2}{V_0^2 + V_1^2} \; .
\end{equation}

\noindent This asymptotic limit is indicated in Figure~\ref{fig:Surf_waves} by the
horizontal dashed lines. We note that this dispersion relation is identical to
that achieved for surface waves residing on the interface between two uniform media
in Cartesian geometry when the transverse wavenumber is large \citep[see ][]{Wentzel:1979}.
This is a strong indication that in this limit, the surface wave becomes insensitive
to the curvature of the field lines and to the stratification on either side of the
interface. The lack of dependence on the stratification is expected because the mode
becomes ever more confined to the interface as the axial wavenumber $k$ increases
(i.e., the mode's skin depth to either side shrinks). The lack of the dependence
on the curvature may be a special property of the potential magnetic equilibrium.
Note that there is an absence of curvature terms appearing in the momentum
equation~\eqnref{eqn:vector_MHD}. Such terms would appear as a radial component of
the tension proportional to $B\Phi/r$. In the derivation of
equation~\eqnref{eqn:vector_MHD}, all such terms were cancelled by corresponding
terms appearing in the magnetic pressure force that arise from the radial variation
of the magnetic field strength.


\acknowledgements

This work was supported by NASA through grants NNX14AG05G, NNX14AC05G, and NNX09AB04G.
We thank the SDO team for the use of AIA/SDO data. RJ would like to acknowledge
the support of the University of Sheffield (UK) and thanks R. Maurya for discussions
regarding AIA/SDO images.




\begin{thebibliography}{}

\bibitem[Andries et al.(2009)]{Andries:2009}
Andries, J., Van Doorsselaere, T., Roberts, B., Verth, G., Verwichte, E.,
\& Erd\'elyi, R. 2009, \ssr, 149, 3

\bibitem[Aschwanden et al.(1999)]{Aschwanden:1999}
Aschwanden M.~J., Fletcher, L., Schrijver, C.~J., \& Alexander, D. 1999, \apj, 520, 880

\bibitem[De Moortel \& Brady(2007)]{DeMoortel:2007}
De Moortel, I. \& Brady, C.S. 2007, \apj, 664, 1210

\bibitem[Dunster(1990)]{Dunster:1990}
Dunster, T.M. 1990, SIAM J. Math. Anal., 21, 995

\bibitem[Gil et al.(2004)]{Gil:2004}
Gil, A., Segura, J., Temme, N.M. 2004, ACM Transactions on Mathematical Software, 30, 159

\bibitem[Hindman \& Jain(2014)]{Hindman:2014}
Hindman, B.~W. \& Jain, R. 2014, \apj, 784, 103

\bibitem[Jain et al.(2015)]{Jain:2015}
Jain, R., Maurya, R.A., Hindman, B. W. 2015, \apj, 804, L19

\bibitem[Nakariakov et al.(1999)]{Nakariakov:1999}
Nakariakov, V., Ofman, L., DeLuca, E., Roberts, B., \& Davila, J.~M. 1999,  Science, 285, 862

\bibitem[Press et al.(2007)]{Press:2007}
Press, W.H., Teukolsky, S.A., Vetterling, W.T., \& Flannery, B.P., 2007,
Numerical Recipes: The Art of Scientific Computing,  Third Edition,
(Cambridge Univ. Press: Cambridge), ISBN-10: 0521880688.

\bibitem[Roberts et al.(1984)]{Roberts:1984}
Roberts, B., Edwin, P.~M., \& Benz, A.~O. 1984, \apj, 279, 857

\bibitem[Schmit \& Gibson(2011)]{Schmit:2011}
Schmit, D.J. \& Gibson, S.E. 2011, \apj, 733, 1

\bibitem[Van Doorsselaere et al.(2007)]{VanDoorsselaere:2007}
Van Doorsselaere, T., Nakariakov, V.M., \& Verwichte, E. 2007, \aap, 473, 959

\bibitem[Verwichte et al.(2009)]{Verwichte:2009}
Verwichte, E., Aschwanden, M.~J., Van Doorsselaere, T., Foullon, C., \& Nakariakov, V.~M. 2009, \apj, 698, 397

\bibitem[Verwichte et al.(2004)]{Verwichte:2004}
Verwichte, E., Nakariakov, V.~M., Ofman, L., \& Deluca, E.~E. 2004, \solphys, 223, 77

\bibitem[Wang \& Solanki(2004)]{Wang:2004}
Wang, T. \& Solanki, S. 2004, \AA, 421, L33

\bibitem[Wentzel(1979)]{Wentzel:1979}
Wentzel, D.G. 1979, \apj, 227, 319

\end{thebibliography}
\end{document}